\documentclass[sigconf, nonacm]{acmart}

\AtBeginDocument{%
  \providecommand\BibTeX{{%
    \normalfont B\kern-0.5em{\scshape i\kern-0.25em b}\kern-0.8em\TeX}}}

\usepackage[utf8]{inputenc}
\begin{document}

\title{Multi-modal Embedding Fusion-based Recommender}


\author{Anna Wróblewska}
\orcid{0000-0002-3407-7570}
\affiliation{%
  \institution{Synerise S.A.}
  \streetaddress{89 Marszałkowska}
  \institution{Warsaw University of Technology}
  \city{Warsaw}
  \state{Poland} 
}
\author{Jacek Dabrowski}
\affiliation{%
  \institution{Synerise S.A.}
}
\email{jack.dabrowski@synerise.com}
\author{Michal  Pastuszak}
\affiliation{%
  \institution{Synerise S.A.}
}
\email{michal.pastuszak@synerise.com}
\author{Andrzej Michalowski}
\affiliation{%
  \institution{Synerise S.A.}
}
\email{andrzej.michalowski@synerise.com}
\author{Michal Daniluk}
\affiliation{%
  \institution{Synerise S.A.}
}
\email{michal.daniluk@synerise.com}
\author{Barbara Rychalska}
\affiliation{%
  \institution{Synerise S.A.}
  \streetaddress{89 Marszałkowska}
  \institution{Warsaw University of Technology}
  \city{Warsaw}
  \state{Poland} 
}
\email{barbara.rychalska@synerise.com}
\author{Mikolaj Wieczorek}
\affiliation{%
  \institution{Synerise S.A.}
}
\email{mikolaj.wieczorek@synerise.com}
\author{Sylwia Sysko-Romanczuk}
\affiliation{%
  \institution{Warsaw University of Technology}
  \city{Warsaw}
  \state{Poland} 
}

\renewcommand{\shortauthors}{Wróblewska and .., et al.}

\begin{abstract}
Recommendation systems have lately been popularized globally, with primary use cases in online interaction systems, with significant focus on e-commerce platforms. We have developed a ML (machine learning)-based recommendation platform, which can be easily applied to almost any items and/or actions domain. Contrary to existing recommendation systems, our platform supports multiple types of interaction data with multiple modalities of metadata natively. This is achieved through multi-modal fusion of various data representations. We deployed the platform into multiple e-commerce stores of different kinds, e.g. food and beverages, shoes, fashion items, telecom operators. Here, we present our system, its flexibility and performance. We also show benchmark results on open datasets, that significantly outperform state-of-the-art prior work.


\end{abstract}



\maketitle

\section{Introduction}
Recommender systems aim to suggest relevant items to users (items being movies to watch, texts to read, products to buy or anything else depending on industries). Indeed, the systems are present at almost every large e-commerce store or platform, spanning various sectors from garments, through jewellery to food.

There exist multiple frameworks and algorithms to build recommender systems, and the choice of the optimal approach strongly depends on the types of data available, the distributional properties of the data, modalities considered and business use cases~\cite{ijcai2019-529,aaai2019-fusing,10.1145/2939672.2939673,hidasi2015sessionbased,10.1145/3132847.3132926}. \textbf{It is usually impossible to adjust existing algorithms to include a new modality of data or a new type of attributes.} Hence, a vast majority of existing recommender systems consider only a single type of interaction, e.g. clicks or purchases - yet even in this simple scenario, the generalization of performance to various datasets seems doubtful~\cite{10.1145/3298689.3347058}.

Businesses using large datasets desire, outside of the currently used data range, a system that will be based on predictors derived from variables generated through automatic analysis of customers' voices (audio) and observations how customers interact with the merchant's websites, as well as mobile and offline ecosystems. Unlike existing solutions, it is expected to use these variables by correlating them in real time with data from other channels, which would significantly increase the systems' effectiveness, as well as expand its functionality. Improvement in effectiveness of customer behavior predictive analytics is a key challenge for many businesses.

We show our innovative recommender system that utilizes a multi-modal fusion of multiple interaction types (e.g. clicks, purchases, adding a product to cart) and multiple attribute modalities (audio, video, images, text, other behavioral data through time). Our system provides a very efficient framework to combine, deploy and evaluate different recommendation algorithms and scenarios utilizing rich, multi-modal and multi-view data sources.

In this work we present our contributions to recommender systems in data science. We define the requirements for a next generation recommender system as follows:
\begin{itemize}
    \item multiple input interaction types (e.g. clicks, purchases, add-to-cart, geo-locations),
    \item multiple input attribute modalities (e.g. text, image, video, other),
    \item ease of adding new back-end algorithms,
    \item effective deep learning models for visual search, recommendations with and without session information, which outperform state-of-the-art techniques,
    \item specialized techniques to fuse multiple modalites, 
    \item high efficiency and scalability (services architecture),
    \item convenient infrastructure for model evaluation and performance measurements.
\end{itemize}

In the following sections we describe our recommendation system architecture and the data workflow ~\ref{sec:over}. Then main features -- such as multi-modal embeddings and their fusion technology -- are sketched in sections~\ref{sec:emb} and \ref{sec:fus}, respectively. Subsequently we present a few tests with state-of-the-art (SotA) benchmarks (section~\ref{sec:bench}). Finally we add a description of our interface, recommendation analytics and a few use-cases from our production deployments (section~\ref{sec:demo}).

\section{Motivation and state-of-the-art review}
\label{sec:mot}

Predictive analytics incorporated in recommender systems unleash the power of data for users and businesses. Systems which learn from data how to predict future choices and behaviors of individuals can bring significant competitive advantages. While perfect prediction is not possible in practice, sufficiently developed systems can bring benefits for both customers and businesses. This paper focuses mainly on the customer side.

The purpose of recommendation is to help users find the products they need, manage their personal budget efficiently and make purchase decisions faster. This is usually achieved by showing related offers and recommending similar products to the ones they have viewed, recommending the next products to consider or to complement a shopping cart. 

There exist multiple established recommendation algorithms, ranging from simple heuristic-based methods (such as KNNs), through Collaborative Filtering to deep learning architectures~\cite{10.1145/3331184.3331210,10.1145/2959100.2959167,DBLP:journals/corr/abs-1811-00855,10.1145/3298689.3347058}.

Different algorithms are useful in different input data settings, use cases and scenarios. Common similar items recommendations based on text and numeric data involves preparing suggestions (i.e. other items or actions to take) considering the context of a single item. Personalized recommendations suggest the products considering the context of users' buying preferences and their behavioral profiles (based on long- or short-term history). In order to prepare these suggestions, the system analyzes page visits, transactional data and also product feeds (product metadata). There are also other types of recommendations, i.e. cross-sell, top products, last seen offers.

In practice the techniques are often mixed depending on environment and various factors, e.g. vendor domain, website construction, user history and current season or time of the day. They should be adjusted experimentally and measured constantly.

\begin{figure*}[!ht]
  \centering
  \includegraphics[width=15cm]{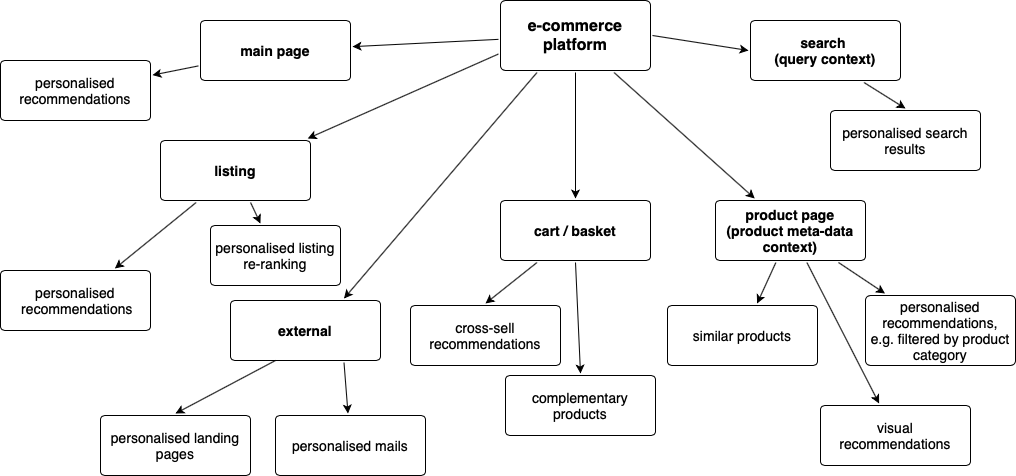}
  \caption{Diagram of various recommendation scenarios depending on the site within an e-commerce platform}
  \label{fig:scenarios}
\end{figure*}

Thus, a system that is highly self-adjustable to the type and modality of data is crucial to cope with many deployments and to use recommendation techniques effectively. A set of recommendation scenarios (which we also use as default settings in our platform) are shown in Fig.~\ref{fig:scenarios}.

\section{Our Approach}

\subsection{System Overview}
\label{sec:over}
\begin{figure*}[!ht]
  \centering
  \includegraphics[width=16cm]{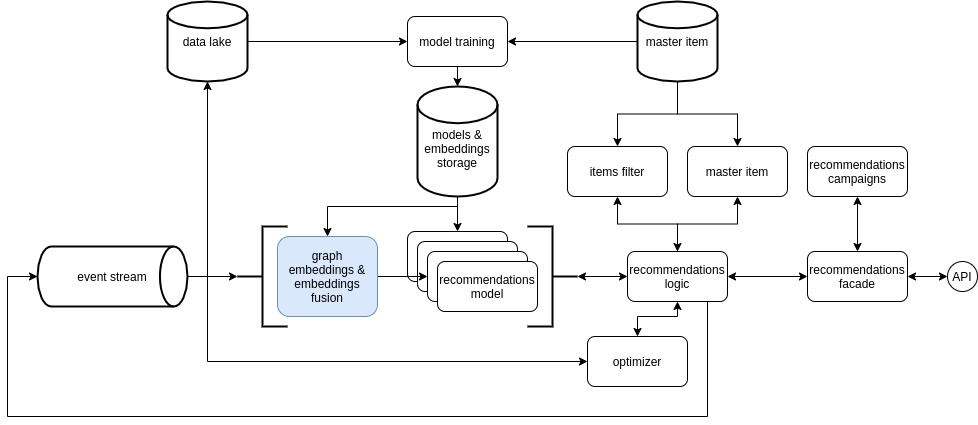}
  \caption{A general architecture of our recommendation platform}
  \label{fig:architecture}
\end{figure*}

Our platform is adjusted to consume business customers' standard format APIs for product feeds and our proprietary product catalog database infrastructure. (see~Fig.~\ref{fig:architecture}).

The system is based on Reactive Microservices Architecture~\cite{RMA_2016,rmanifesto}, implementing its core principles which are: elasticity, scalability, fault tolerance, high availability, message driven and real-time processing. Especially real-time processing is crucial in order to provide tailored and high quality recommendations taking into account not only the latest changes of in-session user behavior, but also changes in system performance. Not only scores and recommendations are being calculated during the request time, but also user representations are being updated and exposed to models after each event flowing through event stream.

The conceptual diagram of an architecture is presented on Fig.~\ref{fig:architecture}. The system is accessible throughout an extensive API which is exposed by recommendations facade. When a new request for recommendation appears, before it is be passed to recommendation logic module, it is validated by the facade and enriched with business rules via recommendation campaigns. Rules may include things like: type of recommendation, recommendation goal or filtering expressions formulated in our dedicated control language, i.e. items query language IQL. 

\begin{figure}[!ht]
  \fbox{\includegraphics[width=8cm]{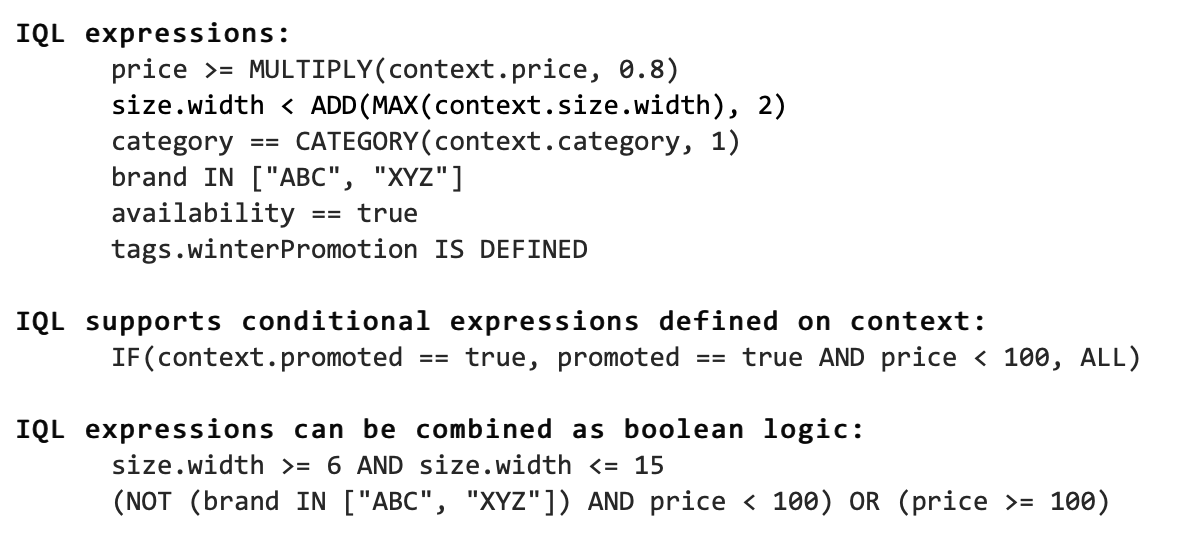}}
  \caption{Items query language implemented and used in our recommendation system}
  \label{fig:iql}
\end{figure}

IQL custom query language provides a very flexible framework to build new recommendation scenarios based on item meta-data and recommendation request context. In Fig.~\ref{fig:iql} there are a few examples of building recommendation filtering rules. IQL expressions are being handled by an items filter, which performs filtering of candidate items based on given constraints. To achieve high throughput and low latency, items filter uses its own compressed binary representation of items, serving thousands of requests per second and filtering sets of million+ items. In case of IQL expressions with low selectivity, transfer of the data structure containing candidate item IDs over the network infrastructure could be expensive, therefore a binary protocol between filter and logic has been implemented. The model which will handle the request is selected by the Optimizer. Optimizer implements a form of a Thompson Sampling algorithm solving multi-armed bandit problems allowing not only to easily A/B test new ideas and algorithms, but also to optimize results of running recommendation campaigns. Finally one of the models receives a request to score available candidates based on model itself and to update entity embeddings.

Although most of the system works in real time, the offline part is also present but limited to model training. Algorithms are trained on two main data sources. The first one is a data lake into which events of different types and origins are being ingested through an events stream. To name a few events types: screen view from a mobile app, product add to cart from a web page, offline transaction from a POS system etc. The second source is a master item meta-data database where items are being kept along with their attributes and rich data types like images.

\subsection{Multi-modal Embeddings}
\label{sec:emb}

Our algorithms can be fed with various kinds of input data. The system analyzes long- and short-term interaction history of users and has a deep insight into item metadata. For this purpose we use a multi step pipeline, starting with unsupervised learning. For images and texts off-the-shelf unsupervised models may be used. For interaction data we identify graphs of user-entity interactions (e.g. user-product, user-brand, user-store) and compute multiple graph or network embeddings.

We developed a custom method\footnote{The module responsible for the method is shown in light blue in our system diagram in Fig.~\ref{fig:architecture}.} for massive-scale network embedding for networks with hundreds of billions of nodes and tens of billions of edges. The task of network embedding is to map a network or a graph into a low-dimensional embedding space, while preserving higher-order proximities between nodes. In our datasets nodes represent interacting entities, e.g. users, device IDs, cookies, products, brands, title words etc. Edges represent interactions, with a single type of interaction per input network, e.g. purchase, view, hover, search.

Similar network embedding approaches include Node2Vec, DeepWalk and RandNE~\cite{zhang2018billion}. These approaches exhibit several undesirable properties, which our method addresses. Thanks to the right design of algorithm and highly optimized implementation our method allows for:
\begin{itemize}
    \item three orders of magnitude improvement in time complexity over Node2Vec and DeepWalk,
    \item deterministic output -- embedding the same network twice results in the same embeddings,
    \item stable output with regards to small input perturbations -- small changes in the dataset result in similar embeddings,
    \item inductive property and dynamic updating -- embeddings for new nodes can be created on the fly,
    \item applicable to both networks and hyper-networks -- support for multi-node edges.
\end{itemize}

The input data is constructed from raw interactions - an edge (hyperedge) list for both simple networks and hypernetworks. In case of hypernetworks, where the cardinality of an edge is larger than 2, our algorithm either performs implicit clique expansion in-memory (to avoid excessive storage needs for an exploded input file). For very wide hyperedges star-expansion results in less edges, and can be used instead - via an input file containing virtual interaction nodes.

Our custom method works as follows: At first we initialize node vectors (Q matrix) randomly via multiple independent hashing of node labels and mapping them to constant interval, resulting in vectors sampled from uniform (-1, 1) distribution. Thus we achieve deterministic sampling. Empirically we determine that dimensionality of 1024 or 2048 is enough for most purposes.
Then we calculate a Markov transition matrix (M) representing network connectivity. In case of hyper-network, we perform clique expansion adding virtual edges.
Final node embeddings are achieved by multiplying $M*Q$ iteratively and L2-normalizing them in each intermediate step. The number of iterations is depends on the distributional properties of the graph, with between 3 and 5 iterations being a good default range.


The algorithm is optimized for extremely large datasets:
\begin{itemize}
    \item The Markov transition matrix M is stored in COO (co-occurrence) format in RAM or in memory-mapped files on disk;
    \item all operations are parallelized with respect to the embedding dimensions, because dimensions of vectors Q are independent on each other;
    \item the $M*Q$ multiplication is performed with dimension-level concurrency as well;
    \item clique expansion for hyper-graphs is performed virtually, only filling the entries in $M$ matrix;
    \item star expansion is performed explicitly, with a transient column for the virtual nodes in the input file.
\end{itemize}

The algorithm's results are entity embeddings contained in the $Q$ matrix. Creation of inductive embeddings (for new nodes) is possible from raw network data using the formula $M'*Q$, where $M'$ represents the links between existing and new nodes and $Q$ represents the embeddings of existing nodes. 

It is worth noting that the algorithm not only performs well on interaction networks, but also on short text data, especially product metadata. In this setting we consider words in a product title as a hyperedge. This corresponds to star-expansion, where product identifiers are virtual nodes linking title words.

However our general pipeline can easily use embeddings calculated using the latest techniques of language modeling, e.g. ELMO, BERT embeddings, especially for longer texts.

Another data source is  visual data (shape, color, style, etc.) i.e. images. To prepare visual data feed for our algorithm
we use state-of-the-art deep learning neural networks \cite{kucer_detect-then-retrieve_2019,dodds_learning_2018} customized for our use~\cite{wieczorek2020strong}. 

Indeed, any unsupervised learning method outputting dense embeddings can be considered as input to our general pipeline.

\subsection{Embedding Fusion}
\label{sec:fus}

Having unsupervised dense representations coming from multiple, possibly different algorithms - representing products, or other entities the customers interacts with, we need to aggregate them into fixed-size behavioral profiles for every user.

As most methods of representation learning assume nothing about embedding compositionality (with simple assumptions made by Bag-of-Words models), we develop a custom mechanism of compositionality allowing meaningful summation of multiple items.

Our algorithm performs multiple feature space partitionings via vector quantization. The algorithm involves ideas derived from Locality Sensitive Hashing and Count-Min Sketch algorithm, combined with geometric intuitions. Sparse representations resulting from this approach exhibit additive compositionality, due to Count-Sketch properties (for a set of items, the sketch of the set is equal to the sum of separate sketches).

All modalities and views of data (all embedding vectors) are processed in this way, their sketches are concatenated.

One of the central advantages of the algorithm is the ability to squash representations of multiple objects into a much smaller joint representation which we call   (\textit{sketch}), which allows for easy and fast subsequent retrieval of participating objects, in an analogous way to Count-Min Sketch. E.g. the purchase history of a user can be represented in a single sketch, the website browsing history as another sketch, and the sketches concatenated.

Subsequently sketches containing squashed user behavioral profiles serve as input to relatively shallow (1-5 layers) feed-forward neural networks. The output structure of the neural network also is structured as a sketch, with the same structure.

Training is done with cross-entropy objective in a depth independent way (output sketches are normalized to 1, across the width dimension). During inference, we perform a sketch readout operation, as in a classic Count-Min Sketch, exchanging the minimum operation to geometric mean - effectively performing averaging of log-probabilities.

\section{A Few Experiments and Results on Open Datasets}
\label{sec:bench}

As far as visual similarity is concerned we tested our proprietary deep learning models on big open datasets commonly used in this field, i.e. DeepFashion and Street2Shop ~\cite{kucer_detect-then-retrieve_2019, kuang_fashion_2019}. Our models are better than SotA in general and in various garment categories as well~\cite{wieczorek2020strong} (see Tab.~\ref{tab:sota_comparison_vs}). 

For history/session based model comparison with SotA we used a framework published in \cite{ludewig2019empirical} and our results are better or comparable to the results of the SotA methods depending on metrics considered (see Tab.~\ref{tab:sota_comparison_sbr}).

Prelimiary results on MovieLens 20M dataset \cite{MovieLens-data} regarding featured recommendations (without user history) show that our proprietary models are comparable to SotA in the field ($P@20$ above 20\%; we used benchmark published in \cite{microsoft_2019}).
Moreover, our algorithm offers significant speed benefits over other neural competitors. It takes 20 sec to train and 14 sec to return predictions for 6000 users and 4000 movies (around 23.000.000 user/movie combinations in total), compared to recent neural approaches: FastAI recommender \cite{howard2018fastai} (901 sec/57 sec) or NCF \cite{he_2017} (790 sec/50 sec) while achieving comparable results \cite{microsoft_2019}, using the same hardware.


\begin{table}[!ht]
\begin{center}
\caption{Comparison of performance on our models and currently published SotA in visual similarity research. Metrics are commonly used in the task~\cite{kucer_detect-then-retrieve_2019}}
\label{tab:sota_comparison_vs}
\begin{tabular}{l|ccc}
         \hline 
         \multicolumn{4}{c}{Dataset - Street2Shop} \\
         Model & mAP  & Acc@1 & Acc@20 \\ 
         \hline
         single model \cite{kucer_detect-then-retrieve_2019} & 26.1 & 29.9  & 57.6 \\
         ensamble model \cite{kucer_detect-then-retrieve_2019} &  29.7 & 34.4  & 60.4 \\
         our custom single model  &  \textbf{37.2}      &  \textbf{42.3}      & \textbf{61.1}   \\
         \hline
         \multicolumn{4}{c}{Dataset - DeepFashion} \\
         Model & Acc@1 & Acc@20 & Acc@50 \\ 
         \hline
         single model \cite{dodds_learning_2018} & 27.5 & 65.3  & 76.0 \\
         our custom single model  &  \textbf{30.8}      &  \textbf{69.4}      & \textbf{78.0}   \\
         \hline 
\end{tabular}
\end{center}
\end{table}

\begin{table}[!ht]
\begin{center}
\caption{Comparison of performance on our models and currently published SotA in session-based recommenders research. Metrics are commonly used in the task~\cite{ludewig2019empirical}}
\label{tab:sota_comparison_sbr}
\begin{tabular}{l|cccccc}
         \hline 
         \multicolumn{4}{c}{Dataset - RETAIL} \\
         Model & MAP@20 & P@20 & R@20 & HR@20 & MRR@20 \\ 
         \hline
         STAN \cite{stan2019} & 0.0285 & 0.0543 & 0.4748 & 0.5938 & 0.3638 \\
         our model  &  \textbf{0.0302} & \textbf{0.0556} & \textbf{0.4974} & \textbf{0.6202} & \textbf{0.3649} \\
         \hline
         \multicolumn{4}{c}{Dataset - DIGI} \\
         Model & MAP@20 & P@20 & R@20 & HR@20 & MRR@20 \\ 
         \hline
         SKNN \cite{Ludewig_2018} & 0.0255 & 0.0596 & 0.3715 & 0.4748 & 0.1714 \\
         VSTAN \cite{Ludewig_2018} & 0.0252 & 0.0588 & 0.3723 & 0.4803 & \textbf{0.1837} \\
         our model & \textbf{0.0268} & \textbf{0.0620} & \textbf{0.3849} & \textbf{0.4908} & 0.1731\\
         \hline 
\end{tabular}
\end{center}
\end{table}

\section{Use cases}
\label{sec:demo}

In current production deployments our platform achieves 20\% - 30\% improvements in average order size (AOS) and 10\% - 60\% improvements in average order value (AOV) in comparison to the system without our custom ML-based recommendations as calculated by A/B tests. The numbers vary significantly depending on quality of product and user data, as well as recommendation visibility and website structure. 

\begin{figure}[!ht]
  \centering
  \includegraphics[width=8cm]{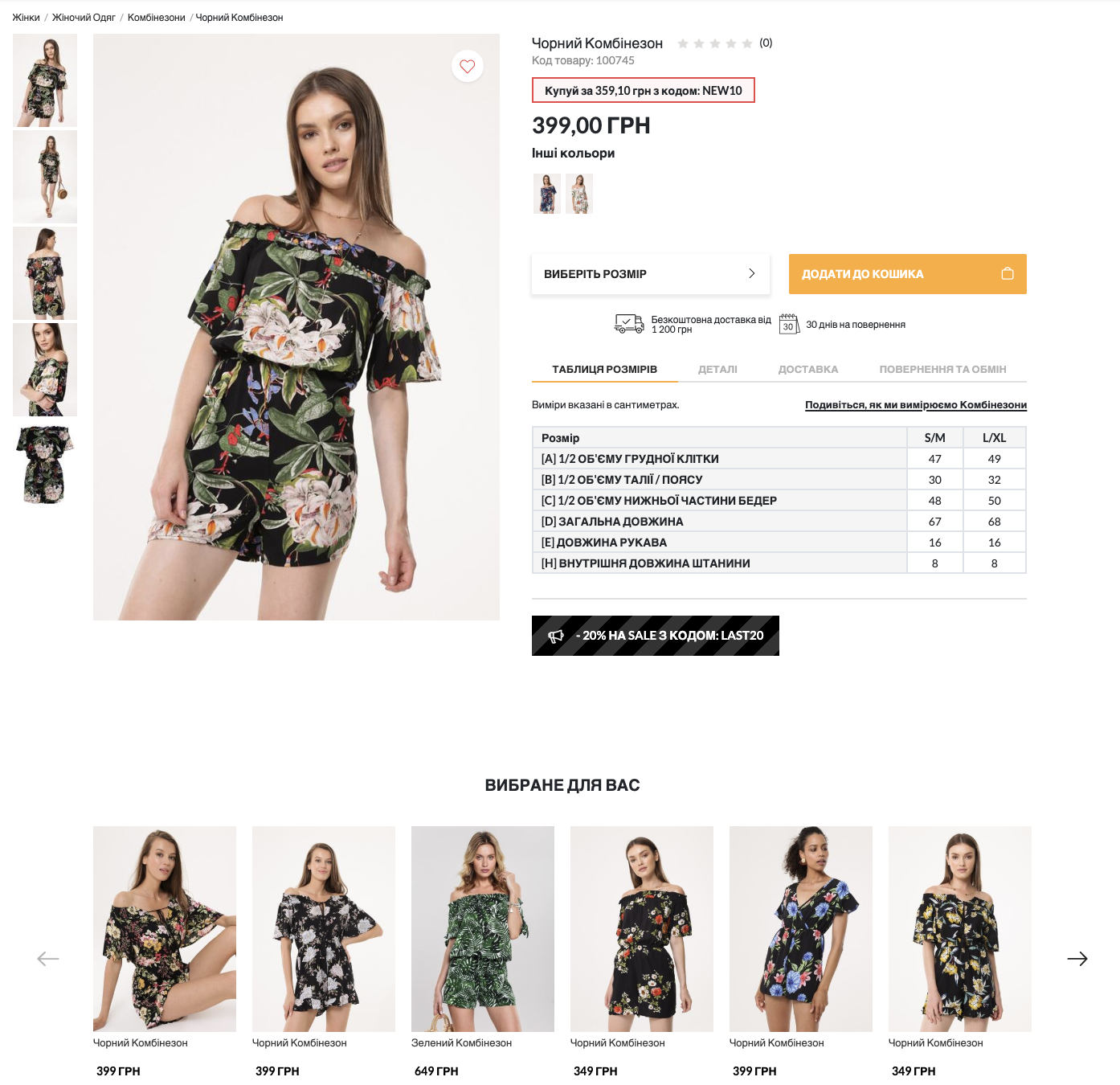}
  \caption{Recommendation box -- visually similar dresses}
  \label{fig:vs-box}
\end{figure}

\begin{figure*}[!ht]
  \centering
  \includegraphics[width=14cm]{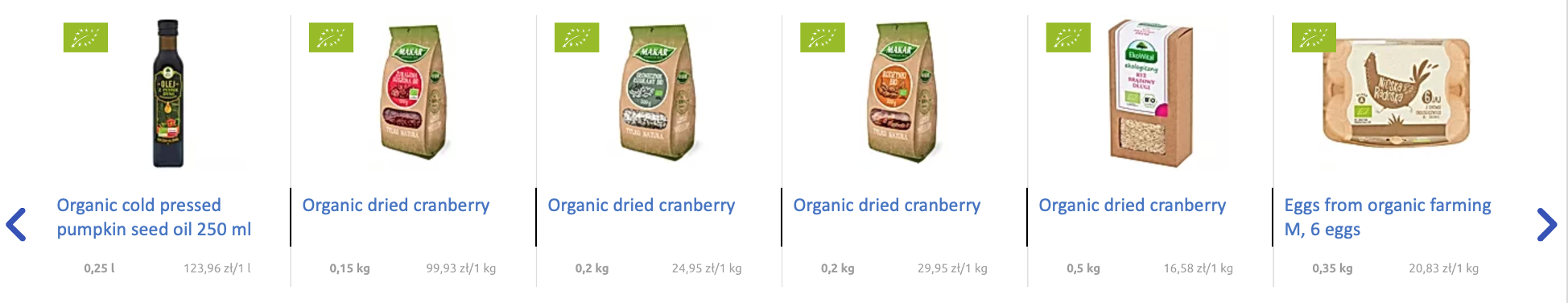}
  \caption{Recommendation box at main page of a service -- personalized on user interactions in the same session }
  \label{fig:food-box}
\end{figure*}

\begin{figure*}[!ht]
  \centering
  \includegraphics[width=14cm]{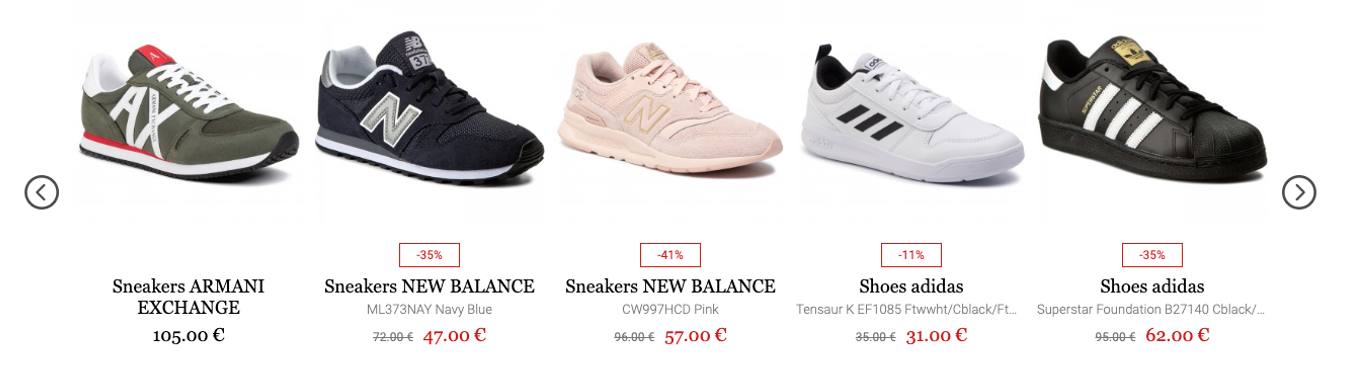}
  \caption{Recommendation box -- cold start in a session based on previous user behavioral history}
  \label{fig:shoe-box}
\end{figure*}

In Fig.~\ref{fig:vs-box},\ref{fig:food-box},\ref{fig:shoe-box} different recommendation scenarios in various product categories are shown, these are visually similar products and personalized recommendations based on user interactions in various e-commerce platforms. 

In Tab.~\ref{tab:behavioral-electronics} we also provide data about user history and recommendations in electronics category.

\begin{table*}[!ht]
\begin{center}
\caption{Examples of recommendations for our personalized recommenders in electronic products. We can also see complementary products that fit very well to the viewed product.}
\label{tab:behavioral-electronics}
\begin{tabular}{l|l}
\hline\hline 
\begin{tabular}{p{8cm}}
\textbf{History:} \\
Page Visit: HUAWEI P20 Lite Smartphone Pink \\
Page Visit: HUAWEI P20 Lite Smartphone Pink \\
\hline
\textbf{User really bought later:}
HUAWEI P20 Lite Smartphone Pink \\
\hline
\textbf{Recommendations:} \\
HUAWEI P20 Lite Smartphone Pink \\
HUAWEI P20 Lite Smartphone Black \\
HUAWEI P20 Lite Blue Smartphone \\
MYSCREEN Lite Edge tempered glass for Huawei P20 Lite Black \\
HUAWEI Transparent case for Huawei P20 Lite Transparent \\
MERCURY Jelly Case for Huawei P20 Lite Transparent \\
HAMA Crystal Clear Cover case for Huawei P20 Lite Transparent \\
HAMMER case for Huawei P20 Lite Black \\
HUAWEI Smart Cover for Huawei P20 Lite Black \\
\hline\hline
\textbf{History:} \\
Transaction: KARCHER RM 500 cleaner for 500 ml glass\\
\hline
\textbf{User really bought later:}
KARCHER WV 5 Premium window cleaner 1.633-453.0 \\
\hline
\textbf{Recommendations:} \\
KARCHER WV 5 Premium window cleaner 1.633-453.0\\
KARCHER WV Classic 1.633-169.0 window washer\\
KARCHER WV 2 Premium window cleaner 1.633-430.0\\
KARCHER RM 500 cleaner for 500 ml glass\\
KARCHER telescopic lance for WV window washer\\
KARCHER cleaner RM 503 500 ml\\
KARCHER microfiber pad for WV window washer (2 pieces)\\
KARCHER cleaner in 500 ml canister (RM561)    \\
\end{tabular} & 
\begin{tabular}{p{8cm}}
\textbf{History:} \\
Page Visit: Smartphone APPLE iPhone 11 Pro Max 256GB Space Gray\\
Page Visit: Fridge HAIER A3FE737CMJ\\
Page Visit: Washing machine AEG LTX7E272P\\
Page Visit: Smartphone APPLE iPhone 11 Pro Max 256GB Space Gray\\
Page Visit: Smartphone APPLE iPhone 11 Pro Max 256GB Night green\\
Page Visit: Smartphone APPLE iPhone 11 Pro Max 256GB Space Gray\\
\hline
\textbf{User really bought later:}
Smartphone APPLE iPhone 11 Pro Max 256GB Star Gray\\
\hline
\textbf{Recommendations:} \\
Smartphone APPLE iPhone 11 Pro Max 256GB Star Gray\\
Smartphone APPLE iPhone 11 Pro Max 256GB Night green\\
Smartphone APPLE iPhone 11 Pro Max 64GB Star gray\\
HOFI Glass Pro + tempered glass for Apple iPhone 11 Pro Max\\
Hybrid glass HOFI Hybrid Glass for Apple iPhone 11 Pro Max Black\\
APPLE Silicone Case for iPhone 11 Pro Max Black\\
APPLE Leather Case for iPhone 11 Pro Max Black\\
SPIGEN Neo Hybrid case for Apple iPhone 11 Pro Max Navy-silver\\
Watch Dogs 2 Game PS4\\
\end{tabular} \\
\hline\hline
\end{tabular}
\end{center}
\end{table*}

\begin{figure}[!ht]
  \centering
  \includegraphics[width=8cm]{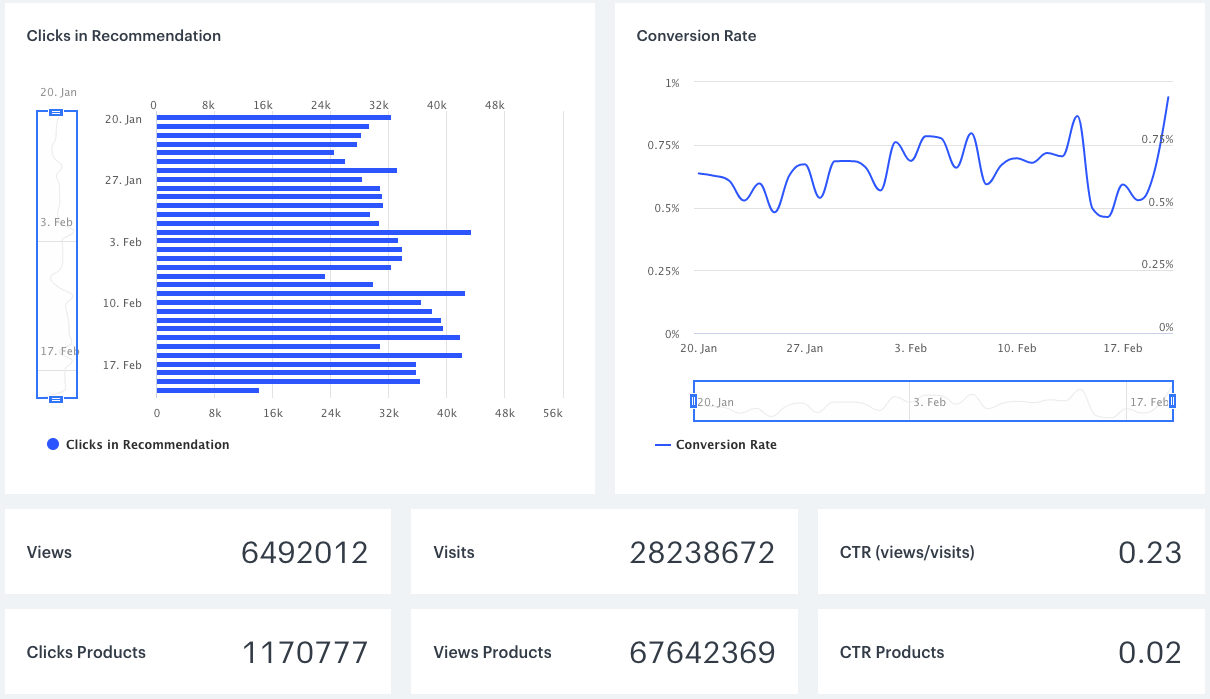}
  \caption{Recommendation analytics interface -- aggregated results}
  \label{fig:vs-analytics}
\end{figure}

\begin{figure}[!ht]
  \centering
  \includegraphics[width=8cm]{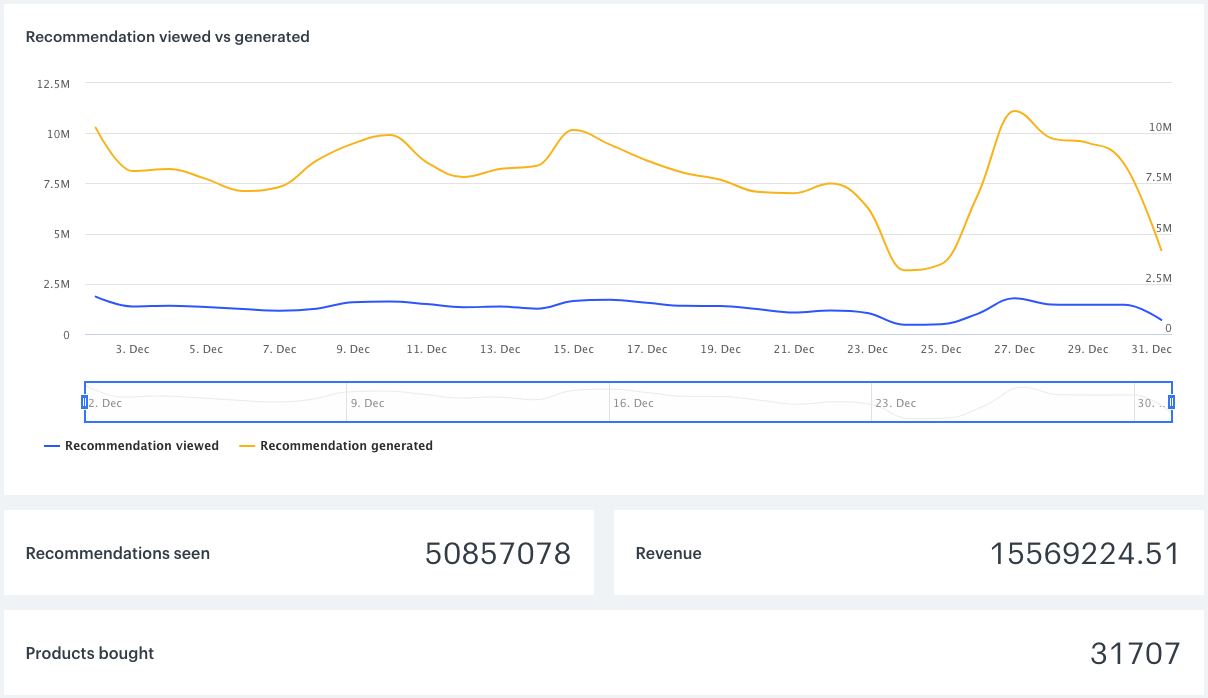}
  \caption{Recommendation analytics plot - viewed and clicked recommenations}
  \label{fig:gen-view-plot}
\end{figure}

Recommendation analytics in our systems provide easily customized interface to show aggregated results and plots (for an example see Fig.~\ref{fig:vs-analytics} and Fig.~\ref{fig:gen-view-plot}). 

\section{Discussion and Conclusion}
In this paper we presented our custom platform and algorithms that can be fed with multimodal, multi-view data, fused and aggregated effectively. We showed use cases with various scenarios and data feeds. With our algorithms, described in previous sections, we achieved very good results in various e-commerce stores and we exceeded state-of-the-art results on open recommendation datasets. Deployment of our system in a new e-commerce store takes about one workday, thanks to a modular architecture easily adaptable to clients' APIs and data feeds of different formats.

Given the elegant nature of sketch representations of any embeddings learned in an unsupervised way, allowing for compact representation and additive composability, future applications are numerous. Our focus for current work includes: product propensity models, demand forecasting, improved search personalization and recommendation of non-product entities (e.g. coupons, offers, brands). We are also extending our framework with interpretability functionality (XAI methods) to be even more useful in business contexts.

\bibliographystyle{ACM-Reference-Format}
\bibliography{sample-base}

\end{document}